\documentclass[aps,prl,twocolumn,superscriptaddress,showpacs,reprint]{revtex4-1}

\usepackage{graphicx}
\usepackage{hyperref} 
\usepackage[usenames]{xcolor}
\usepackage{amsmath}

\newcommand{\be}{\begin{equation}}
\newcommand{\ee}{\end{equation}}
\newcommand{\ba}{\begin{eqnarray}}
\newcommand{\ea}{\end{eqnarray}}

\def\simge{\mathrel{\rlap{\raise 0.5ex
     \hbox{$>$}}{\lower 0.5ex \hbox{$\sim$}}}}
\def\simle{\mathrel{\rlap{\raise 0.5ex
      \hbox{$<$}}{\lower 0.5ex \hbox{$\sim$}}}}

\newcommand{\Singlet}{\mbox{$^1$S$_0$}}

\newcommand{\fig}[1]{Fig.~\ref{#1}}

\begin{document} 
%%%%%%%%%%%%%%%%%%%%%%%%%%%%%%%%%%%%%%%%%%%%%%%%%%%%%%%%%%%
\title{Forecasting neutron star temperatures: predictability and variability} 
\author{\surname{Dany} Page}
\affiliation{Instituto de Astronom\'{\i}a, 
                  Universidad Nacional Aut\'onoma de M\'exico, 
                 Mexico D.F. 04510, Mexico}
\author{\surname{Sanjay} Reddy}
\affiliation{Institute for Nuclear Theory, 
                  University of Washington, 
                  Seattle, Washington 98195, USA}

%%%%%%%%%%%%%%%%%%%%%%%%%%%%%%%%%%%%%%%%%%%%%%%%%%%%%%%%%%%
\begin{abstract} 
It is now possible to model thermal relaxation of neutron stars after bouts of accretion during which the star is heated 
out of equilibrium by nuclear reactions in its crust.
Major uncertainties in these models can be encapsulated in modest variations of a handful of fudge parameters
that change the crustal thermal conductivity, specific heat, and heating rates. 
Observations of thermal relaxation constrain these fudge parameters and 
allow us to predict longer term variability in terms of the neutron star core temperature. 
We demonstrate this explicitly by modeling ongoing thermal relaxation in the neutron star XTE J1701-462.
Its future cooling, over the next 5 to 30 years, is strongly constrained and depends mostly on its core temperature, 
uncertainties in crust physics having essentially been pinned down by fitting to the first three years of observations.   
\end{abstract} 
%%%%%%%%%%%%%%%%%%%%%%%%%%%%%%%%%%%%%%%%%%%%%%%%%%%%%%%%%%%

% These were up-date when uploading the paper to the PRL site:
% 26.60.Gj		(Neutron star crust)
% 97.60.Jd		(Neutron Stars)
% 97.80.Jp		(X-ray binaries)
% 98.70.Qy		(X-ray sources; X-ray bursts)

% Previous pacs: PRL upload site did not like 26.60.+c
% 97.60.Jd   Neutron stars
% 95.30.Cq  Elementary particle processes
% 26.60.+c   Nuclear matter aspects of neutron stars 

\maketitle 

%%%%%%%%%%%%%%%%%%%%%%%%%%%%%%%%%%%%%%%%%%%%%%%%%%%%%%%%%
%%%%%%%%%%%%%%%%%%%%%%%%%%%%%%%%%%%%%%%%%%%%%%%%%%%%%%%%%

%            I N T R O D U C T I O N

%%%%%%%%%%%%%%%%%%%%%%%%%%%%%%%%%%%%%%%%%%%%%%%%%%%%%%%%%
%%%%%%%%%%%%%%%%%%%%%%%%%%%%%%%%%%%%%%%%%%%%%%%%%%%%%%%%%

Neutron star-low mass x-ray binaries (NS-LMXBs) are close binary systems in which a neutron star is accreting matter from a low-mass companion 
through Roche lobe overflow \cite{Van-der-Klis:2006fk}.
In many systems accretion is not continuous, and during quiescence, when presumably no, or very little, accretion is occurring, thermal radiation from the neutron star surface can be observed to infer its temperature, $T_e$ \cite{Rutledge:1999uq}.
In recent years, five NS-LMXBs were observed to return to quiescence after a long accretion outburst
that lasted years (1.3 to 24) and have been called quasi-persistent transients (QPTs)
\cite{Cackett:2010fk,Cackett:2008uq,Fridriksson:2010fk}.
Recent attempts to model the observed evolution of $T_e$ after the end of an accretion outburst
have revealed that its temporal characteristics are likely set by processes in the outer $1-2$ km region 
of the neutron star called the crust  \cite{Shternin:2007ly,Brown:2009zr,Page:2012ys}. 

During accretion, the gravitational energy released at the surface is radiated away but non-equilibrium reactions occurring deep inside the crust heat the neutron star interior. 
These reactions arise as accretion induced compression of matter drives electron captures, 
neutron emissions and absorptions, and pycno-nuclear (i.e., induced by pressure) fusion reactions \cite{Bisnovatyi-Kogan:1979ve,Haensel:1990bh}.
Together, these processes liberate about $1.5 - 2$ MeV of energy per accreted baryon in the density interval $10^9-10^{14}$ g cm$^{-3}$ 
\cite{Gupta:2008xw,Haensel:2008vn}. 
This process, called deep crustal heating \cite{Brown:1998dq}, gradually heats the whole neutron star interior that will
reach a stationary temperature $T_0$ determined by the long term average mass accretion rate,
$\langle \dot M \rangle$, and the core neutrino luminosity \cite{Colpi:2001fk}.

On a shorter time scale, characteristic of QPTs, heating during a bout of accretion is strong enough to drive the crust out of 
thermal equilibrium with the core \cite{Rutledge:2002cr} and 
cooling observed immediately after should reveal the thermal relaxation of the neutron star crust.
Theoretical modeling of the two systems KS 1731-260 and MXB 1659-29 (``KS'' and ``MXB'' hereafter) 
 has confirmed this expectation and  shown QPTs are unique laboratories for studying neutron star crust physics
 \cite{Shternin:2007ly,Brown:2009zr}.
Although matter in the crust is at high density ($\rho \simeq 10^{9}-10^{14}$ g cm$^{-3}$) where novel quantum and superfluid behavior is expected, nuclear interactions are fairly well understood at these sub-nuclear densities and a theoretical framework to describe the structure and thermal properties of matter exists \cite{Pethick:1995di,Page:2012ys}. Models of the neutron star crust are now sufficiently advanced that key uncertainties associated with the thermal conductivity, 
heat capacity and nuclear reactions rates needed to describe thermal relaxation have been identified and studied \cite{Chamel:2008}. 
More importantly, as we shall discuss later, together theory and observation already constrain the expected range of variation in these quantities. 

In this letter, we combine a detailed theory of the crust with previous constraints and early time cooling data from a specific source XTE J1701-462 (``XTE'' hereafter)
to make predictions for the future evolution of its surface temperature. 
Our predictions can be tested in the near term and we find that further cooling is very strongly correlated with its core temperature. 
In the long term, as additional sources entering a transient cooling phase are discovered, they will further test theoretical predictions and 
establish the basic paradigm that observed cooling is due to thermal and transport phenomena in the crust. Our numerical simulations use the code {\em NSCool} \cite{NSCool}, an updated version of the code used in \cite{Colpi:2001fk}.

%%%%%%%%%%%%%%%%%%%%%%%%%%%%%%%%%%%%%%%%%%%%%%%%%%%%%%%%%
%%%%%%%%%%%%%%%%%%%%%%%%%%%%%%%%%%%%%%%%%%%%%%%%%%%%%%%%%

%            C R U S T   P H Y S I C S

%%%%%%%%%%%%%%%%%%%%%%%%%%%%%%%%%%%%%%%%%%%%%%%%%%%%%%%%%
%%%%%%%%%%%%%%%%%%%%%%%%%%%%%%%%%%%%%%%%%%%%%%%%%%%%%%%%%

%++++++++++++++++++++++++++++++++++++++++++++++++++++++++++++++++++++++
\begin{figure*}[floatfix]
   \begin{center}
   \includegraphics[width=0.99\textwidth]{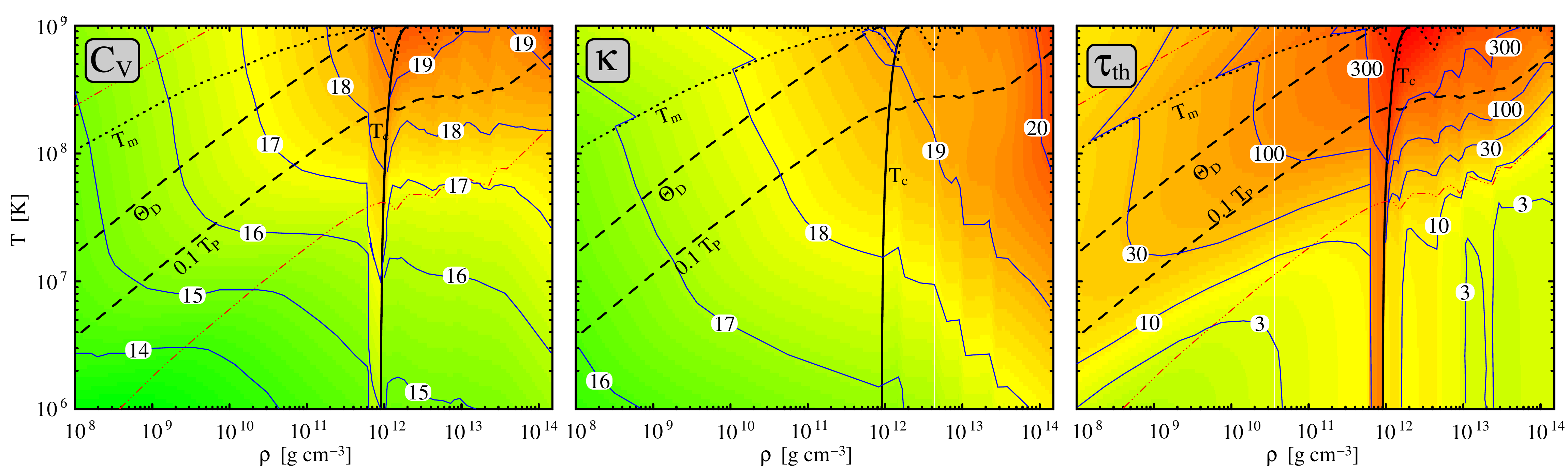}
  \end{center}
   \caption{The ``neutron star crust landscape''.
   Left panel: color plot of the specific heat $C_V$, in erg K$^{-1}$cm$^{-3}$, with (blue) contour lines 
   labelled by $\mathrm{log}_{10} C_V$.
   Central panel: color plot of thermal conductivity $\kappa$, in erg K$^{-1}$cm$^{-1}$s$^{-1}$, with (blue) contour lines
   labelled by $\mathrm{log}_{10} \kappa$.
   Right panel: color plot of $\tau_\mathrm{th} = C_V/\kappa$ in time/(length)$^2$ with (blue) contour lines at
   3, 10, 30, 100, and 300 days per (100 m)$^2$.
   In all three panels $Q_\mathrm{imp}^{\mathrm lo \,\rho } = 20$  at $\rho$ below $10^{12}$ g cm$^{-3}$ and
   $Q_\mathrm{imp}^{\mathrm hi \,\rho } = 4$ above $10^{13}$ g cm$^{-3}$, with a smooth transition inbetween.
   The neutron $\Singlet$ gap is from  \cite{Schwenk:2003fk} which has a
   layer of unpaired neutron only just above neutron drip: its $T_c$ is shown.
   Also plotted on each panel is the ion melting temperature, $T_\mathrm{m}$ \cite{Slattery:1982tg}, 
   the Debye temperature $\Theta_D \simeq 0.45 T_P$ \cite{Carr:1961ys},
   and $0.1 \, T_\mathrm{P}$, $T_\mathrm{P}$ being the ion plasma temperature.
   In the $C_V$ and $\tau_\mathrm{th}$ panels, the two (red) dash-triple dot lines demarks the boundaries between which
   $C_V$ is dominated by the ions.
    }
   \label{Fig1} 
\end{figure*}
%++++++++++++++++++++++++++++++++++++++++++++++++++++++++++++++++++++++

Evolution of temperature in the crust is determined by the heat diffusion equation 
\be
C_V \frac{\partial T}{\partial t} = \kappa \frac{\partial^2 T}{\partial r^2} + 
\frac{1}{r^2} \frac{\partial (r^2 \kappa)}{\partial r} \frac{\partial T}{\partial r}+ Q_h - Q_\nu
\label{eq:master}
\ee
where $C_V$ is the specific heat, $\kappa$ the thermal conductivity, and the other factors, denoted  by $ Q_h $ and  $Q_\nu $, are the nuclear heating and the neutrino cooling rates, respectively.  These density and temperature dependent inputs have been studied in some detail.  These dependencies are now well fairly understood (for a recent review see  \cite{Page:2012ys}) and major uncertainties can parameterized in terms of a handful of density dependent effective parameters: 
(i) the local effective ion-plasma  frequency $\omega_{\rm P}=(4\pi Z^2e^2~n_{\rm ion}/M_{\rm ion}^*)^{1/2}$, 
where  $n_{\rm ion}$ is the total ion density and $M_{\rm ion}^*$ is the ion effective mass  which incorporates effects due to entrainment in the inner crust \cite{Chamel:2012ix},
$Ze$ being the ion electric charge; 
(ii) the transition temperature $T_{\rm c}$ for neutron superfluidity in the inner crust
and (iii) the impurity parameter $Q_\mathrm{imp} = \sum_{i}~n_i(Z_i -  \langle Z \rangle )^2/n_{\rm ion}$ 
where $n_i$ is the number density of the impurity species "$i$" of charge $Z_i e$, $\langle Z \rangle e$
being the average ion charge.
Nuclear reactions that generate heat have been also studied and here we employ the heating rates from electron capture and pycno-nuclear reactions from \cite{Haensel:2008vn,Gupta:2008xw}. Despite large uncertainties in the pycno-nuclear reaction rates, the net heating is rather well constrained by global energetics  \cite{Haensel:2008vn}.  

Due to its high thermal conductivity the core temperature remains nearly uniform and its evolution is slow due to the high specific heat. Consequently, the thermal time-scale is largely set by the crust. 
A  simples estimate 
%obtained by neglecting variations of the ratio $C_V/\kappa$ 
gives 
\be
\tau_\mathrm{th} \sim  \frac{C_V }{\kappa} ~(\Delta r)^2
\label{Eq:tau_th}
\ee
where $\Delta r$ is the thickness of the evolving layer. An example of the variation of $C_V$, $\kappa$ and the ratio $ C_V/\kappa$ are shown in the left, middle and right panels of Fig.~\ref{Fig1}, respectively. We will now briefly discuss the main sources of uncertainty for $C_V$ and $\kappa$ and  the range of their variability that we will employ in our simulations. 

%%%%%%%%%%%%%%%%%%%%%%%%%%%%%%%%%%%%%%%%%%%%%%%%%%%%%%%%%
%%%%%%%%%%%%%%%%%%%%%%%%%%%%%%%%%%%%%%%%%%%%%%%%%%%%%%%%%

%            S P E C I F I C   H E A T

%%%%%%%%%%%%%%%%%%%%%%%%%%%%%%%%%%%%%%%%%%%%%%%%%%%%%%%%%
%%%%%%%%%%%%%%%%%%%%%%%%%%%%%%%%%%%%%%%%%%%%%%%%%%%%%%%%%

The ions form a quantum Coulomb crystal at low temperature when $T < 0.1~ T_\mathrm{P}$ where $T_\mathrm{P} = \hbar \omega_\mathrm{P}/k_B$ is the ion plasma temperature.
At these low temperatures, the ion component specific heat is dominated by the transverse phonons contribution
and $C_V^\mathrm{ion} \propto T^3/v_t^3$ where 
$v_t \propto \omega_{\rm P}/q_{\rm D}$ is the transverse phonon velocity and $q_{\rm D}=(6\pi^2~n_\mathrm{ion})^{1/3}$ the Debye momentum. 
The velocity of phonon modes remains somewhat uncertain because coupling between 
dynamics of the neutron superfluid and the lattice is not known precisely and results in significant variation of $C_V^\mathrm{ion}$ \cite{Kobyakov:2013zr,Chamel:2012ix}.
We incorporate this uncertainty by considering, at each depth in the inner crust, a range of possible ion effective mass $M^*_\mathrm{ion}$, and hence a range of $\omega_{\rm P}$,
from $A m_n$ to $A_\mathrm{cell} m_n$, where $A_\mathrm{cell} = A + A_\mathrm{drip}$, $A$ being the ion mass number
and $A_\mathrm{drip}$ the number of dripped neutron in the Wigner-Seitz cell.
When $\Theta_\mathrm{D} \simle T \simle T_\mathrm{m}$, $T_\mathrm{m}$ being the ion melting temperature, the ion specific heat $C_V^\mathrm{ion}$ is almost $T$-independent and $\simeq 3k_B$
while it slowly decreases in the liquid phase.
In contrast, the electron component is well approximated by a degenerate ultra-relativistic Fermi-Dirac gas so that the electron specific heat is simply $C^\mathrm{e}_V=T p_{\rm Fe}^2 (k_B^2/3\hbar^2 c)$, where $p_{\rm Fe}$ is the electron Fermi momentum. For $T \ll T_\mathrm{P}$ electrons dominate since $C_V^\mathrm{e} \propto T$ but with increasing $T$ ions take over, since
$C_V^\mathrm{ion} \propto T^3$.
In the inner crust, neutrons are superfluid below a critical temperature denoted by $T_c$ and their contribution to $C_V$ is suppressed by the factor $\propto \exp{(-T/T_c)}$ for $T \ll T_c$, 
and somewhat enhanced near $T\simeq T_c$ \cite{Levenfish:1994hc}. 
Generically, in a thin layer from the neutron drip point $\rho_\mathrm{drip} \simeq 6\times 10^{11}$  g cm$^{-3}$ up to $\sim 10^{12}$ g cm$^{-3}$, 
where $T \simge T_c$, the neutron contribution is large and accounts for the $C_V$ barrier seen in the left and right panels of \fig{Fig1}.

%%%%%%%%%%%%%%%%%%%%%%%%%%%%%%%%%%%%%%%%%%%%%%%%%%%%%%%%%
%%%%%%%%%%%%%%%%%%%%%%%%%%%%%%%%%%%%%%%%%%%%%%%%%%%%%%%%%

%            T H E R M A L   C O N D U C T I V I T Y

%%%%%%%%%%%%%%%%%%%%%%%%%%%%%%%%%%%%%%%%%%%%%%%%%%%%%%%%%
%%%%%%%%%%%%%%%%%%%%%%%%%%%%%%%%%%%%%%%%%%%%%%%%%%%%%%%%%

Electrons dominate thermal conduction and their contribution is given by $\kappa_{\rm e} = C^\mathrm{e}_V~c^2/(3~\nu^{e})$ where $\nu^\mathrm{e}$ is the electron scattering rate.
When $T\lesssim T_\mathrm{P}$ the scattering rate is dominated by impurity scattering, even for relatively small values of the impurity parameter $Q_\mathrm{imp} \sim 1$
The electron-impurity scattering rate is given by   
\begin{equation}
\nu^\mathrm{e}_{\rm imp} =\nu^\mathrm{e}_0~\frac{Q_{\rm imp}}{\langle Z^2 \rangle}~\Lambda_{\rm imp} 
\end{equation} 
where $ \nu^\mathrm{e}_0 = 4 \alpha_{\rm em}^2~\langle Z^2 \rangle~p_{\rm Fe}/(3\pi~\langle Z) \rangle ) $ is the fiducial electron  scattering rate in an uncorrelated gas, and 
$\Lambda_{\rm imp} \approx 2$ 
is the Coulomb logarithm for randomly distributed impurities  \cite{Flowers:1976}. The elastic impurity scattering rate is independent of temperature and relatively insensitive to lattice and superfluid dynamics.  

At shallow depth $Q_\mathrm{imp}$ is set by the nuclear reactions at the surface where explosive burning through the r-p process can produce highly impure mix with $Q_\mathrm{imp}\simeq 30-100$ \cite{Schatz:2001xx}. However,  several processes including chemical separation due to preferential freezing of large Z elements at the bottom of the ocean, and neutron-rearrangement and pycnonuclear reactions deeper in the inner crust are expected to greatly reduce $Q_\mathrm{imp}$ in the solid regions of the crust \cite{Horowitz:2007,Gupta:2008xw,Steiner:2012}. 
We treat $Q_\mathrm{imp}(\rho)$ as a free parameter and allow a density dependence,
with a value $Q_\mathrm{imp}^\mathrm{hi \, \rho}$, expected to be small, at high densites and $Q_\mathrm{imp}^\mathrm{lo \, \rho}$, that could be much higher, at low densities.
Modeling crust relaxation in KS and MXB has shown that $Q_\mathrm{imp}^\mathrm{hi \, \rho} \simeq1 -5$  is necessary \cite{Brown:2009zr}.

%%%%%%%%%%%%%%%%%%%%%%%%%%%%%%%%%%%%%%%%%%%%%%%%%%%%%%%%%
%%%%%%%%%%%%%%%%%%%%%%%%%%%%%%%%%%%%%%%%%%%%%%%%%%%%%%%%%

%            D E S C R I P T I O N   O F   T H E   T H R E E   G U Y S

%%%%%%%%%%%%%%%%%%%%%%%%%%%%%%%%%%%%%%%%%%%%%%%%%%%%%%%%%
%%%%%%%%%%%%%%%%%%%%%%%%%%%%%%%%%%%%%%%%%%%%%%%%%%%%%%%%%

%++++++++++++++++++++++++++++++++++++++++++++++++++++++++++++++++++++++T
\begin{figure}[t]
   \begin{center}
   \includegraphics[width=0.48\textwidth]{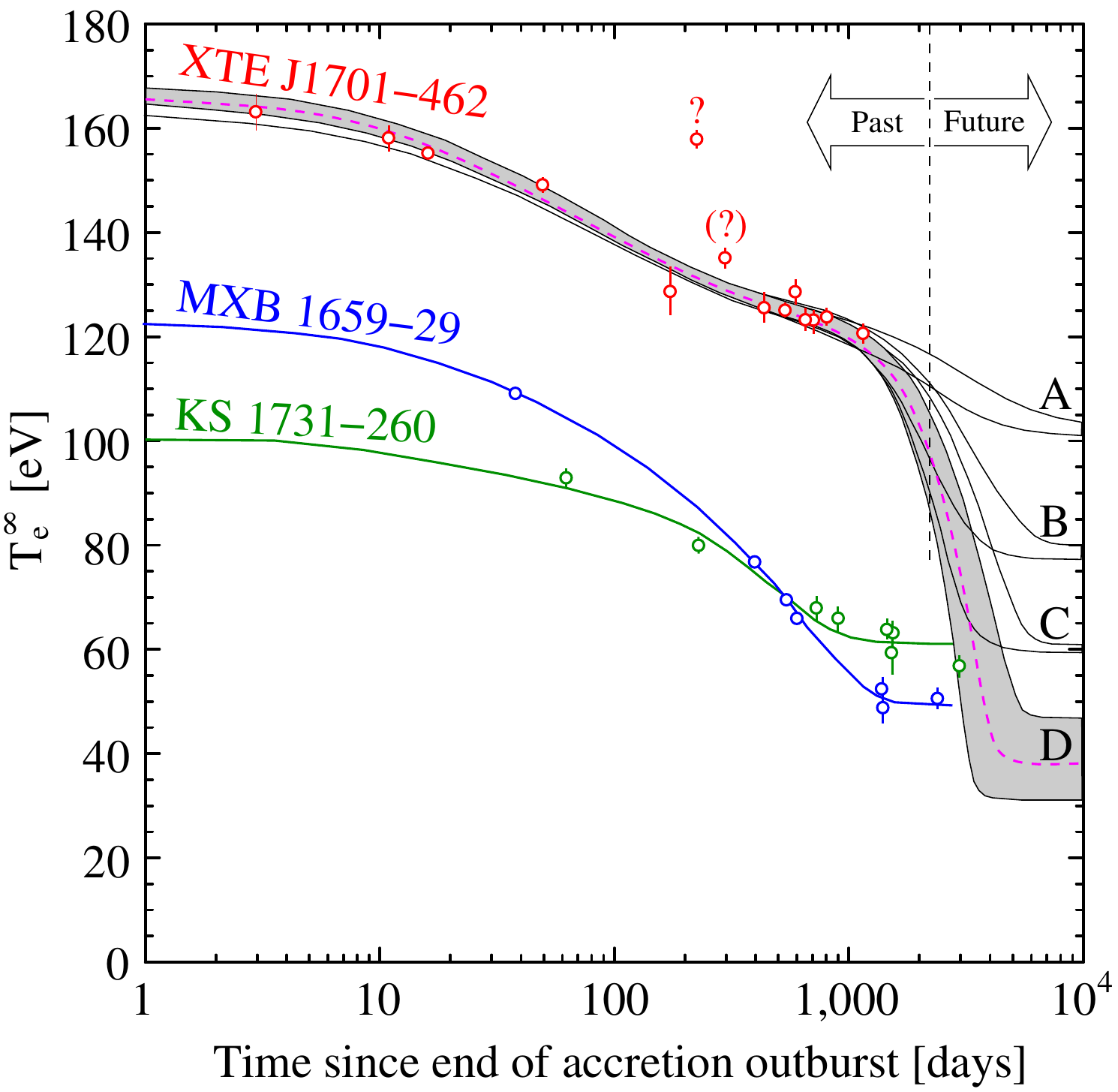}
  \end{center}
   \caption{Observed effective temperatures at infinity, $T_e^\infty$ (circles with $1\sigma$ error bars), after the end
   of the accretion outburst of KS 1731-260 \cite{Cackett:2010fk}, MXB 1659-29 \cite{Cackett:2008uq}, and 
   XTE J1701-462 \cite{Fridriksson:2011kx}.
   Our theoretical models for KS and MXB, similar to those of \cite{Brown:2009zr}, employ the crust physics displayed in \fig{Fig1}
   but with $(Q_\mathrm{imp}^{\mathrm lo \,\rho}, Q_\mathrm{imp}^{\mathrm hi \,\rho})$ equal to  
   (5,3) for KS and (10,3) for MXB \cite{Thermo}.
   The dashed (magenta) curve model for XTE uses exactly the physics of \fig{Fig1} \cite{Thermo} 
   and details are displayed in \fig{Fig2}.
   The two data point marked as ``?''  and ``(?)'' are likely and possibly, respectively, contaminated by residual accretion \cite{Fridriksson:2011kx}.
   All our models that fit the 12 data points of XTE, up to the last one at 1158 days but excluding the two marked with ``?'', 
   are shown in four bands according to the initial core temperature:
    A: $T_0 = 10^8$ K; B: $T_0 = 10^{7.75}$ K; C: $T_0 = 10^{7.5}$ K, and 
    D that comprises all our models with $T_0$ between $10^{7.25}$ and $10^6$ K.
    The ``Past'' and ``Future'' refer to XTE's present time.
   }
   \label{Fig3} 
\end{figure}
%++++++++++++++++++++++++++++++++++++++++++++++++++++++++++++++++++++++

Among the five known QPTs, XTE  is a peculiar system in that during its 1.6 year long outburst 
it accreted at a high rate close to the Eddington limit, $\dot M_\mathrm{Edd} \simeq 2 \times 10^{-8} \, M_\odot$ yr$^{-1}$
\cite{Fridriksson:2011kx}.
For comparison, MXB  and KS had accretion outburst that lasted 2.5 and 12.5 years with average $\dot M$
of $\sim 0.05 \dot M_\mathrm{Edd}$ and $\sim 0.2 \dot M_\mathrm{Edd}$,  respectively
\cite{Cackett:2008uq,Fridriksson:2011kx} .
The observed cooling light-curves of these three stars are displayed in \fig{Fig3}.
XTE's evolution is characterized by a short initial cooling phase of a few hundred days followed by a two year long plateau.
This is in sharp contrast to the observed cooling behaviors of KS and MXB that initially evolved on longer time scales
of about $\sim 300$ and $\sim 450$ days
followed by a very slow evolution \cite{Cackett:2010fk,Cackett:2013fk}. Moreover, XTE  has, on average, a twice larger $T_e^\infty$
than MXB and KS, implying that its crust temperature is about four times higher.
XTE's evolution explores a new, hotter, regime of the neutron star crust landscape of \fig{Fig1} which,
together with its high $\langle \dot M \rangle$ and short outburst, explains its peculiar behavior as we describe below.

%%%%%%%%%%%%%%%%%%%%%%%%%%%%%%%%%%%%%%%%%%%%%%%%%%%%%%%%%
%%%%%%%%%%%%%%%%%%%%%%%%%%%%%%%%%%%%%%%%%%%%%%%%%%%%%%%%%

%            J 1 7 0 1   H E A T I N G   A N D   C O O L I N G

%%%%%%%%%%%%%%%%%%%%%%%%%%%%%%%%%%%%%%%%%%%%%%%%%%%%%%%%%
%%%%%%%%%%%%%%%%%%%%%%%%%%%%%%%%%%%%%%%%%%%%%%%%%%%%%%%%%

%++++++++++++++++++++++++++++++++++++++++++++++++++++++++++++++++++++++
\begin{figure*}[t]
   \begin{center}
   \includegraphics[width=0.45\textwidth]{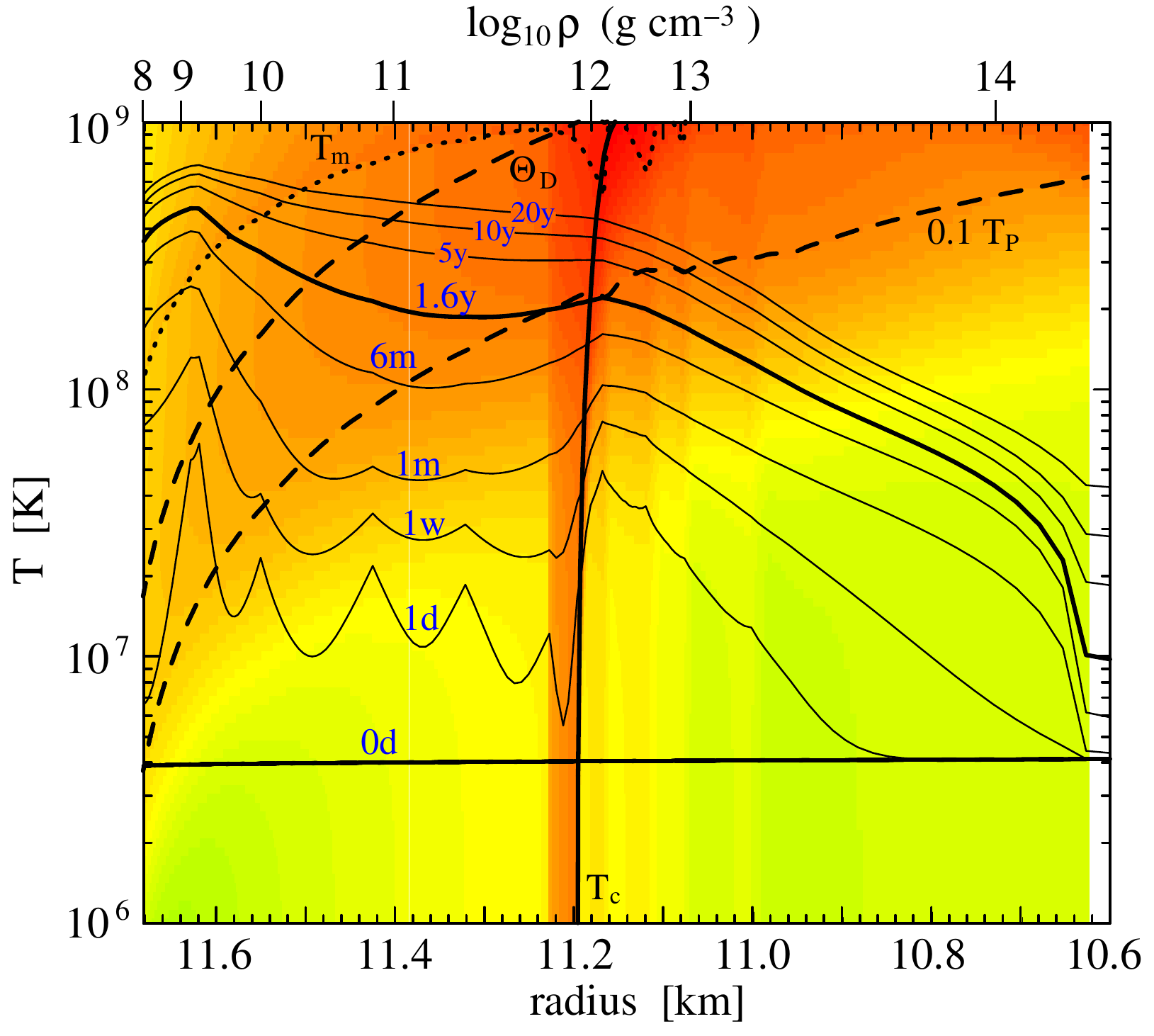}
   \includegraphics[width=0.45\textwidth]{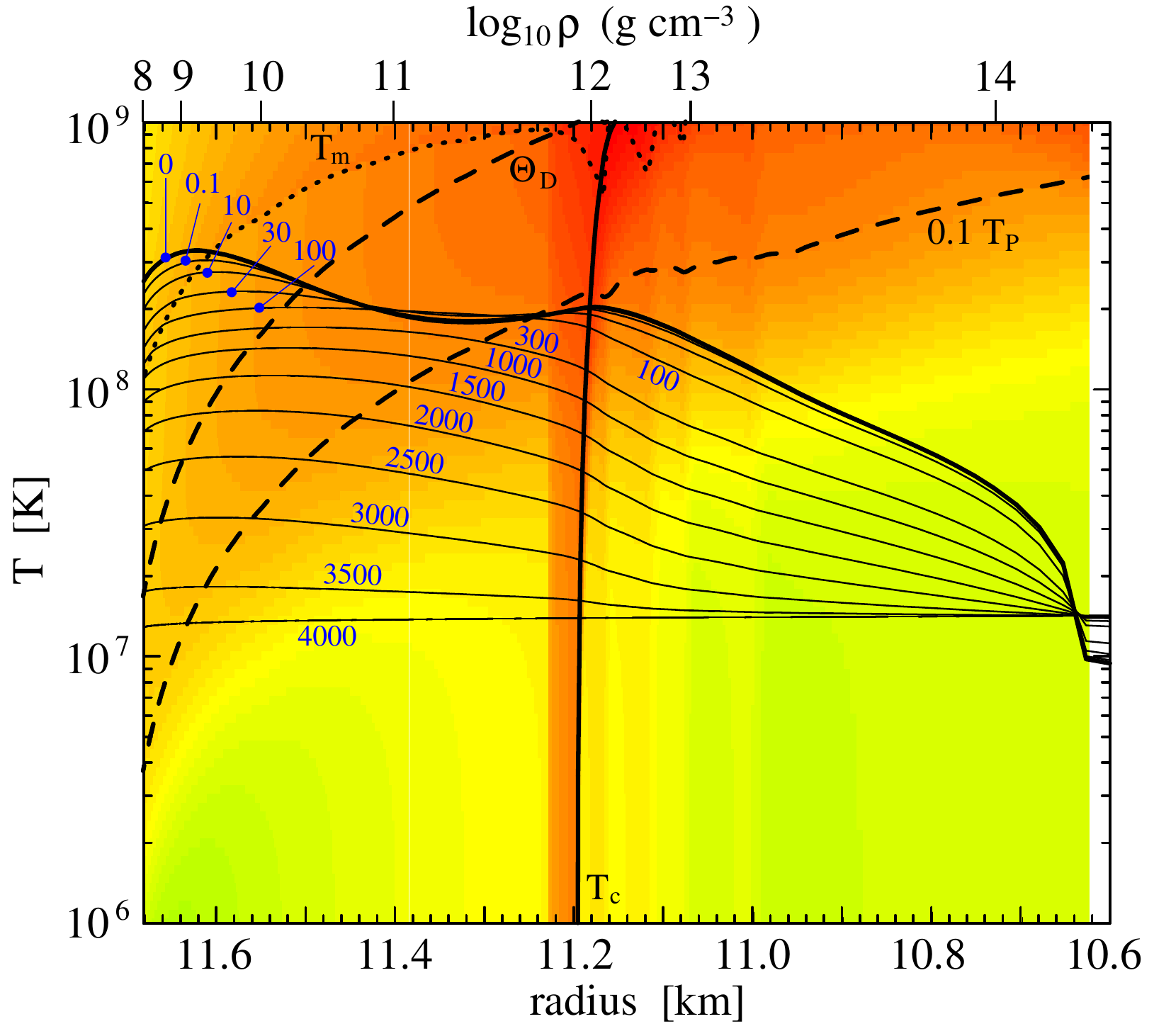}
  \end{center}
   \caption{Left panel: an example of evolution of XTE's crust temperature during an accretion phase at $\dot M \simeq 0.9 \dot M_\mathrm{Edd}$ 
   with initial uniform $T = 4\times 10^6$ K (labelled ``0d'').  Profiles after 1 day (``1d''), 1 week (``1w''), 1 and 6 months
   (``1m'' and ``6m'') and 1.6 year (``1.6y'') are shown. 
   The profiles at 5, 10 and 20 years (``5y'', ``10y'' and ``20y'') show the times needed to approach the steady state.
   Right panel: An example of the evolution of XTE's crust temperature profile during the cooling phase. 
   In both panels the background color map is the local thermal time from Fig.~\ref{Fig1}.
   Notice that the core temperature is increasing both during the accretion phase and the subsequent relaxation phase.
   }
   \label{Fig2} 
\end{figure*}
%++++++++++++++++++++++++++++++++++++++++++++++++++++++++++++++++++++++

We show in \fig{Fig2}, left panel, a series of model crust temperature profiles during accretion induced heating for XTE.
The crucial point is that heating was so strong that it would have taken several decades of accretion for XTE to reach a stationary state.
A stationary state is reached when the inverted $T$ gradient in the crust is large enough that heat flow into the core exactly balances deep crustal heating. 
KS and MXB, with lower $\langle \dot M \rangle$, could reach a stationary state during their longer outbursts \cite{Brown:2009zr}.
Crust microphysics displayed in \fig{Fig1} provides a natural explanation for this diversity - the hotter crust in XTE (were $T \simge 10^8$ K) 
has larger thermal timescales than those encountered in KS and MXB where $T <10^8$ K. 

The thermal relaxation of XTE after its outburst is illustrated in the right panel of \fig{Fig2}.
One see that the outermost, initially hot, 200 meter thick layer relaxes rapidly, in about 100 days,
which results in the observed initial rapid decrease of $T_e^\infty$.
This time scale roughly matches the thermal time scale $\tau_\mathrm{th} \sim 30 - 100$ days of this layer at $T \sim 2 \times10^8$ K 
seen in the right panel of \fig{Fig1}. Subsequent temperature evolution is slow:
heat from the outer crust, whose temperature determines the observed $T_e^\infty$, has to flow into the inner crust, and then into the core.
It has to pass through the bottleneck just above neutron drip where $\tau_\mathrm{th}$ is $\sim$ 1 year at $T \simeq 10^8$ K,
and then diffuse several hundreds meters down to the core.
This process explains the existence of a plateau in the observed $T_e^\infty$.
However, on a longer time scale beyond presently published observations, further decrease of $T_e^\infty$ is naturally expected.

%%%%%%%%%%%%%%%%%%%%%%%%%%%%%%%%%%%%%%%%%%%%%%%%%%%%%%%%%
%%%%%%%%%%%%%%%%%%%%%%%%%%%%%%%%%%%%%%%%%%%%%%%%%%%%%%%%%

%            F O R E C A S T I N G

%%%%%%%%%%%%%%%%%%%%%%%%%%%%%%%%%%%%%%%%%%%%%%%%%%%%%%%%%
%%%%%%%%%%%%%%%%%%%%%%%%%%%%%%%%%%%%%%%%%%%%%%%%%%%%%%%%%

To explore how much three years of observations constrain the properties of the neutron star crust and the future
evolution of XTE we performed an extensive search of the parameter space.
 A comprehensive analysis of our results will presented in a forthcoming paper but a synopsis is displayed in \fig{Fig3}
 as four bands of cooling trajectories labelled as ``A'' to ``D''.
The dominant unconstrained parameter is the core temperature and \fig{Fig3} separates all our models that give a fit
to the data with a $\chi^2$ better than 12 in four classes according to their value of $T_0$.
In the case $T_0$ is smaller than $10^8$ K, fitting the model parameters to the 3 years of observed evolution provides
strong constraints and all models in the cases B, C, and D, have crust microphysics very similar to the one depicted in \fig{Fig1}.
(The same microphysics is also compatible with modeling of KS and MXB as shown in \fig{Fig3}.)
The future evolution of XTE appears to be mostly determined by its previous core temperature $T_0$ and, for a given $T_0$,
uncertainty in future time is smaller than a factor of two. 

%%%%%%%%%%%%%%%%%%%%%%%%%%%%%%%%%%%%%%%%%%%%%%%%%%%%%%%%%
%%%%%%%%%%%%%%%%%%%%%%%%%%%%%%%%%%%%%%%%%%%%%%%%%%%%%%%%%

%            C O N C L U S I O N S

%%%%%%%%%%%%%%%%%%%%%%%%%%%%%%%%%%%%%%%%%%%%%%%%%%%%%%%%%
%%%%%%%%%%%%%%%%%%%%%%%%%%%%%%%%%%%%%%%%%%%%%%%%%%%%%%%%%

It is remarkable that the crust relaxation model is able to describe vastly different temporal behavior observed in the  
three sources XTE J1701-462, MXB 1659-29 and KS 1731-260 with very similar input physics in the inner crust. 
It provides a natural explanation for the rapid early cooling observed in XTE
and predicts future cooling solely in terms of one unknown parameter - the core temperature. 
A robust prediction of the crust cooling model is the correlation between the final temperature  
and the future cooling rate. 
Continued monitoring of XTE will be able to test our prediction.
If confirmed it  would firmly establish the crust relaxation as the underlying process, and  
taken together fits to these three sources will provide useful constraints for the thermal and transport properties of the neutron star crust. 
Finally, results displayed in \fig{Fig2} show that even the core response is not negligible, and these systems may open a new window for studying matter at even larger densities.   We hope that the results presented here will motivate a long term program to discover and monitor accreting neutron stars 
with existing and next generation instruments.       
        
%%%%%%%%%%%%%%%%%%%%%%%%%%%%%%%%%%%%%%%%%%%%%%%%%%%%%%%%%

\begin{acknowledgments}
We thank Bob Rutledge for useful discussions at an early stage of this work 
and Andrew Steiner and Joel Fridriksson for comments on this manuscript.
D.P.'s work is partially supported by grants from the UNAM-DGAPA
(\# IN113211) and Conacyt (CB-2009-01, \#132400).
D.P. acknowledges the hospitality of the Theoretical Division at the
Los Alamos National Laboratory, where part of this work was developed.  
The work of S.R. was supported by the DOE Grant No. DE-FG02-00ER41132 and by the 
Topical Collaboration to study \it{Neutrinos and nucleosynthesis in hot and dense matter}.       
\end{acknowledgments}

%%%%%%%%%%%%%%%%%%%%%%%%%%%%%%%%%%%%%%%%%%%%%%%%%%%%%%%%%%%
%%%%%%%%%%%%%%%%%%%%%%%%%%%%%%%%%%%%%%%%%%%%%%%%%%%%%%%%%%%

%\bibliography{Forecasting_NS}

%%%%%%%%%%%%%%%%%%%%%%%%%%%%%%%%%%%%%%%%%%%%%%%%%%%%%%%%%%%%

%%%%%%%%%%%%%%%%%%%%%%%%%%%%%%%%%%%%%%%%%%%%%%%%%%%%%%%%%%%%
%%%%%%%%%%%%%%%%%%%%%%%%%%%%%%%%%%%%%%%%%%%%%%%%%%%%%%%%%%%%
%
%%%%%%%%%%%%%%%%%%%%%%%%%%%%%%%%%%%%%%%%%%%%%%%%%%%%%%%%%%%%
%%%%%%%%%%%%%%%%%%%%%%%%%%%%%%%%%%%%%%%%%%%%%%%%%%%%%%%%%%%%

%%%%%%%%%%%%%%%%%%%%%%%%%%%%%%%%%%%%%%%%%%%%%%%%%%%%%%%%%%%
\end{document}